\documentstyle[12pt]{article}

\catcode`\@=11
\long\def\@makefntext#1{
\protect\noindent \hbox to 3.2pt {\hskip-.9pt
$^{{\ninerm\@thefnmark}}$\hfil}#1\hfill}		

\def\@makefnmark{\hbox to 0pt{$^{\@thefnmark}$\hss}}  

\def\ps@myheadings{\let\@mkboth\@gobbletwo
\def\@oddhead{\hbox{}
\rightmark\hfil\ninerm\thepage}
\def\@oddfoot{}\def\@evenhead{\ninerm\thepage\hfil
\leftmark\hbox{}}\def\@evenfoot{}
\def\sectionmark##1{}\def\subsectionmark##1{}}

\renewcommand{\thefootnote}{\fnsymbol{footnote}}

\newcounter{sectionc}\newcounter{subsectionc}\newcounter{subsubsectionc}
\renewcommand{\section}[1] {\vspace*{0.6cm}\addtocounter{sectionc}{1}
\setcounter{subsectionc}{0}\setcounter{subsubsectionc}{0}\noindent
	{\normalsize\bf\thesectionc. #1}\par\vspace*{0.4cm}}
\renewcommand{\subsection}[1] {\vspace*{0.6cm}\addtocounter{subsectionc}{1}
	\setcounter{subsubsectionc}{0}\noindent
	{\normalsize\it\thesectionc.\thesubsectionc. #1}\par\vspace*{0.4cm}}
\renewcommand{\subsubsection}[1]
{\vspace*{0.6cm}\addtocounter{subsubsectionc}{1}
	\noindent {\normalsize\rm\thesectionc.\thesubsectionc.\thesubsubsectionc.
	#1}\par\vspace*{0.4cm}}

\newcounter{appendixc}
\newcounter{subappendixc}[appendixc]
\newcounter{subsubappendixc}[subappendixc]

\renewcommand{\appendix}[1] {\vspace*{0.6cm}
        \refstepcounter{appendixc}
        \setcounter{figure}{0}
        \setcounter{table}{0}
        \setcounter{equation}{0}
        \renewcommand{\thefigure}{\Alph{appendixc}.\arabic{figure}}
        \renewcommand{\thetable}{\Alph{appendixc}.\arabic{table}}
        \renewcommand{\theappendixc}{\Alph{appendixc}}
        \renewcommand{\theequation}{\Alph{appendixc}.\arabic{equation}}
        \noindent{\bf Appendix \theappendixc #1}\par\vspace*{0.4cm}}

\def\abstracts#1{{

\centering{\begin{minipage}{12.2truecm}\footnotesize\baselineskip=12pt\noindent
	\centerline{\footnotesize ABSTRACT}\vspace*{0.3cm}
	\parindent=0pt #1
	\end{minipage}}\par}}

\renewenvironment{thebibliography}[1]
	{\begin{list}{\arabic{enumi}.}
	{\usecounter{enumi}\setlength{\parsep}{0pt}
\setlength{\leftmargin 1.25cm}{\rightmargin 0pt}
	 \setlength{\itemsep}{0pt} \settowidth
	{\labelwidth}{#1.}\sloppy}}{\end{list}}

\newcounter{itemlistc}
\newcounter{romanlistc}
\newcounter{alphlistc}
\newcounter{arabiclistc}

\newcommand{\fcaption}[1]{
        \refstepcounter{figure}
        \setbox\@tempboxa = \hbox{\footnotesize Fig.~\thefigure. #1}
        \ifdim \wd\@tempboxa > 6in
           {\begin{center}
        \parbox{6in}{\footnotesize\baselineskip=12pt Fig.~\thefigure. #1}
            \end{center}}
        \else
             {\begin{center}
             {\footnotesize Fig.~\thefigure. #1}
              \end{center}}
        \fi}

\newcommand{\tcaption}[1]{
        \refstepcounter{table}
        \setbox\@tempboxa = \hbox{\footnotesize Table~\thetable. #1}
        \ifdim \wd\@tempboxa > 6in
           {\begin{center}
        \parbox{6in}{\footnotesize\baselineskip=12pt Table~\thetable. #1}
            \end{center}}
        \else
             {\begin{center}
             {\footnotesize Table~\thetable. #1}
              \end{center}}
        \fi}

\def\@citex[#1]#2{\if@filesw\immediate\write\@auxout
	{\string\citation{#2}}\fi
\def\@citea{}\@cite{\@for\@citeb:=#2\do
	{\@citea\def\@citea{,}\@ifundefined
	{b@\@citeb}{{\bf ?}\@warning
	{Citation `\@citeb' on page \thepage \space undefined}}
	{\csname b@\@citeb\endcsname}}}{#1}}

\newif\if@cghi
\def\cite{\@cghitrue\@ifnextchar [{\@tempswatrue
	\@citex}{\@tempswafalse\@citex[]}}
\def\citelow{\@cghifalse\@ifnextchar [{\@tempswatrue
	\@citex}{\@tempswafalse\@citex[]}}
\def\@cite#1#2{{$\null^{#1}$\if@tempswa\typeout
	{IJCGA warning: optional citation argument
	ignored: `#2'} \fi}}

 1
 1
 1

\font\ninerm=cmr9

\begin{document}
\begin{flushright}
	 NUHEP-TH-95-14 \\
	 hep-ph/9512249 \\
\end{flushright}
\bigskip
\bigskip

\centerline{\normalsize\bf Hyperfine Splittings in Heavy Quarkonia}
\baselineskip=16pt
\centerline{\normalsize\bf From a New Improved Spin--Dependent Potential
  \footnote{Presented at the International Symposium on Heavy Flavor
  and Electroweak Theory, Beijing, 16-19 August 1995.}}

\vspace*{0.6cm}
\centerline{\footnotesize Robert J. Oakes}
\baselineskip=13pt
\centerline{\footnotesize\it Department of Physics and Astronomy}
\centerline{\footnotesize\it Northwestern University }
\centerline{\footnotesize\it Evanston, IL 60208 USA}
\centerline{\footnotesize E-mail: oakes@fnalv.fnal.gov}
\baselineskip=12pt

\vspace*{0.9cm}
\abstracts{
A formula for an improved spin--dependent potential between a heavy quark
and a heavy antiquark was developed using Heavy Quark Effective Theory
techniques. The leading logarithmic quark mass terms emerging
from loop contributions were explicitly extracted and summed up. There is
no renormalization scale dependence in this new formula and it includes
both the Eichten--Feinberg formula as well as the one--loop QCD result as
special cases. The hyperfine splittings were calculated using the spin--spin
part of this improved formula. For charmonium the $J/\psi$-$\eta_c$ $S$-wave
splitting agrees well with the data while the energy difference $\Delta M_P$
between the center of gravity of the $1^3P_{0,1,2}$ states and the $1^1P_1$
state has the correct sign but is somewhat larger than the experimental
value; however, there are also several other contributions to $\Delta M_P$
of comparable magnitude ($\sim 1$ MeV),  which we discuss.
The corresponding mass differences for the $b\bar{b}$ and $b\bar{c}$ mesons
are also predicted.}

\vspace*{0.6cm}
\normalsize\baselineskip=15pt
\setcounter{footnote}{0}
\renewcommand{\thefootnote}{\alph{footnote}}

Recently, significant progress has been made in the theoretical study of the
spin--dependent potential between a heavy quark and a heavy antiquark.
An improved formula for the spin--dependent potential was developed~\cite{0}
and used to calculate the hyperfine splittings in the $c\bar{c}$, $b\bar{b}$,
and $b\bar{c}$ systems~\cite{-1}. The improved spin dependent potentials were
derived from QCD first principles using the techniques of the Heavy Quark
Effective Theory (HQET)~\cite{18}. The spin--dependent potential was
separated into short distance parts involving  Wilson coefficients and
long distance parts which were expressed in terms of  gauge invariant
correlation functions of the color-electric and  color-magnetic fields
weighted by the Wilson loop path integral~\cite{1}. If the tree level
values for the Wilson coefficients are used the potential reduces to
Eichten's and Feinberg's result~\cite{1}. And using the one--loop values
of the Wilson coefficients, also calculating the correlation functions
to one--loop in perturbation theory, the spin--dependent potential at the
one--loop level in perturbative QCD~\cite{4a,4b} is recovered. However,
the leading logarithmic terms appearing in perturbative calculations
were also summed up in Ref.~[1] using the Renormalization Group
Equation (RGE) to obtain a scale independent result. Therefore, the
spin--dependent heavy quark--antiquark potential derived  in Ref.~[1]
is scale--independent and thus improves upon and generalizes both
Eichten's and Feinberg's result~\cite{1} and the one--loop perturbative
result~\cite{4a,4b}. In addition, this improved result~\cite{0} satisfies
all the general relations among the different parts of the spin--dependent
potential~\cite{2}.

 The general formula for the spin--spin part of the
renormalization--group--improved spin--dependent potential that was derived
in Ref.~[1] was used to calculate the hyperfine spin splittings in the
$c\bar{c}$, $b\bar{b}$, and $c\bar{b}$ systems in Ref.~[2]. Since the
spin--spin potential is a short distance feature, perturbation theory can
reliably be used in the calculation. The result for the $1^3S_1-1^1S_0$
splitting between the $J/\psi$ and the $\eta_c$ agrees well with the
experimental value~\cite{19} and the predictions for
the mass differences $\psi'-\eta_c'$, $\Upsilon(1S)-\eta_b$,
$\Upsilon(2S)-\eta_b'$, and $B_c^*-B_c$ are reasonable. However, the
contribution to the $P$-wave energy difference, $\Delta M_p$, between the
center of the gravity of $1^3P_{0,1,2}$ states and the $1^1P_1$ state,
while having the correct sign, is somewhat larger than the experimental
data~\cite{19} . That is, when the contributions of the leading logarithmic
terms
are summed up and included, the agreement with that data is not as good
as when only the one--loop perturbative spin--spin potential, in which the
leading logarithmic contributions are not summed up and included, is used.
There are several other contributions to the rather
small energy difference $\Delta M_p$ ($\sim 1$ MeV)
which  estimates indicate are of the same order of magnitude as the
spin--spin contribution. It therefore appears that the agreement of
the one--loop perturbative result with the data is probably fortuitous.

 In the derivation~\cite{0} the
renormalized two--particle effective Lagrangian was first calculated to
order $1/m^2$. Then, treating the terms of higher order in $1/m$ in the
effective Lagrangian as perturbations, the four point Green's function
on the Wilson loop~\cite{24} with the time interval $T$ was calculated
in the limit where $m\to \infty$ first followed by $T\to \infty$~\cite{1,22}.
In this limit, using  standard perturbative methods, the large $T$
behavior of the Green's function is of the form
\begin{eqnarray}
I &\propto & e^{-T\epsilon(m,r)}.
\label{en}
\end{eqnarray}
{}From Eq.~(\ref{en}) $\epsilon(m,r)$, the potential energy between the
quark and the antiquark, can be extracted. Expanding $\epsilon(m,r)$ in
powers of $1/m$ each of the spin--dependent potentials can be factorized
into a short distance part, involving Wilson coefficients, and a long
distance part, which can be expressed in terms of correlation functions
of the color-electric and color-magnetic fields weighted by the Wilson--loop
integral. Using the notation of Ref.~[1], the resulting  spin--spin
potential is
\begin{eqnarray}
\Delta H_{ss}(m_1,m_2,r) =
\displaystyle\frac{{\rm\bf{S}_1\cdot {S}_2}}{3m_1m_2}
\left[\;c_3(\mu,m_1)c_3(\mu,m_2)V_4(\mu,r)
-6N_c g_s^2(\mu) d(\mu)\delta^3({\rm\bf r})\right]
\label{hss}
\end{eqnarray}
where $m_1$, $m_2$, and ${\bf S}_1$, ${\bf S}_2$ are the masses and the
spins of the heavy quark and the antiquark, respectively, $\mu$ is the
renormalization subtraction point, $N_c$ is the number of colors, and
$g_s(\mu)$ is the running coupling constant. The Wilson coefficients
$c_3(\mu,m)$ and  $d(\mu)$ were calculated in leading logarithmic
approximation in Ref.~[11] and Ref.~[1], respectively, and are
\begin{eqnarray}
c_3(\mu,m) &=&
\left(\displaystyle\frac{\alpha_s(\mu)}{\alpha_s(m_1)}\right)^{-{9\over 25}},
\label{c3}
\end{eqnarray}
and
\begin{eqnarray}
d(\mu) &=& \displaystyle\frac{N_c^2-1}{8N_c^2}  c_3(m_2,m_1)[1-c_3^2(\mu,m_2)]
\nonumber \\ &=&\displaystyle\frac{N_c^2-1}{8N_c^2}
\left(\displaystyle\frac{\alpha_s(m_2)}{\alpha_s(m_1)}\right)^{-{9\over 25}}
\left[ 1-\left(\displaystyle\frac{\alpha_s(\mu)}{\alpha_s(m_2)}\right)^
{-{18\over 25}} \right].
\label{dmu}
\end{eqnarray}
In Eq.~(\ref{hss}) $V_4(\mu,r)$ is the color magnetic--magnetic correlation
function which can be expressed as
\begin{eqnarray}
V_4(\mu,r) &\equiv &
\lim_{T\to \infty}\displaystyle\int^{T/2}_{-T/2}dz
\int^{T/2}_{-T/2}dz' \frac{g_s^2(\mu) }{T}
\langle B^i({\rm\bf x}_1,z)B^i({\rm\bf x}_2,z')\rangle /\langle
1\rangle ,
\label{v4}
\end{eqnarray}
where $\langle \cdots\rangle$ is defined by
\begin{equation}
\langle \cdots\rangle \equiv \int [dA^\mu]Tr \left\{ P\left[\exp
\left(ig\oint_{C(r,T)} dz_\mu A^\mu(z) \right)
\cdots\right]\right\}_{x\in C} \exp (iS_{YM}(A)).
\label{corr}
\end{equation}
Here $C(r,T)$ represents the Wilson loop~\cite{24}, $P$ denotes the
path ordering, and $r\equiv |{\rm\bf x}_1- {\rm\bf x}_2|$.

We emphasize that this is a general result for the hyperfine part of the
spin--dependent potential to order $1/m^2$.  It absorbs the short distance
contributions to the potential into the coefficients $c_3(\mu,m)$ and
$d(\mu)$ while the long distance contributions to the potential are
contained in the correlation function $V_4(\mu,r)$. Moreover, the result
is independent of the factorization scale since the $\mu$--dependence in
the coefficients cancels the $\mu$--dependence in the correlation function.
The first term in the bracket in Eq.~(\ref{hss}) is a nonlocal term while
the second term is a local one which is generated by mixing with the first
(nonlocal) term under renormalization. We note that if the coefficients are
evaluated at tree level; i.e., $c_3(\mu,m)=1$ and $d(\mu)=0$, the potential
reduces to the Eichten--Feinberg result~\cite{1}. And if these coefficients
are expanded to order $\alpha_s(\mu)$ and the correlation function is also
evaluated only to one--loop, the logarithmic terms in Eq.~(\ref{hss}) then
reduce to the one--loop spin--spin potential~\cite{4a,4b}. Therefore, this
renormalization--group improved potential, Eq.~(2), extends both Eichten's and
Feinberg's result as well as the one--loop perturbative
potential, containing each of these results as special cases.

To first order perturbation theory in $\Delta H_{ss}$ the energy shift
caused by $\Delta H_{ss}(r)$ is
\begin{eqnarray}
\Delta E &=& \displaystyle \int d^3{\rm\bf r} \Psi^*_{l,l_z}({\rm\bf r} )
\Delta H_{ss} (r) \Psi_{l,l_z}({\rm\bf r})
\end{eqnarray}
where $\Psi_{l,l_z}({\rm\bf r}) $ is the nonrelativistic wavefunction of
the bound state with total angular momentum $l$ and $z$-component $l_z$.
For simplicity we suppress spin and color indices  and retain only the
space--dependent indices. Separating the radial part, $u(r)$, we write
$\Psi_{l,l_z}({\rm\bf r}) $ as
\begin{eqnarray}
\Psi_{l,l_z}({\rm\bf r}) &=& u(r)\; Y_{l,l_z}(\theta,\phi),
\end{eqnarray}
where $Y_{l,m}(\theta,\phi)$ are the standard spherical harmonics.

 The radial wavefunction was obtained by numerically solving
the Schr\"{o}dinger equation. For comparison, we used three popular
potential models. One was the Cornell model~\cite{30},
the second one was the logarithmic potential~\cite{31},
and the third one was the improved QCD--motivated potential~\cite{8}.

The calculation also required an expression for the running coupling
constant $\alpha_s(q)$. The familiar RGE, one--loop result is
\begin{eqnarray}
\alpha_s(q) &=&
\displaystyle\frac{4\pi}{b_0\ln\displaystyle
  \frac{q^2}{\Lambda^2_{\overline{\rm MS}}}},
\label{al1}
\end{eqnarray}
where $b_0=11N_c-2N_f$ and $N_f$ is the number of quark flavors. It is clear
from Eq.~(\ref{al1}) that $\alpha_s(q)$ contains a Landau singularity in
the nonperturbative region when $q^2=\Lambda^2_{\overline{\rm MS}}$ and
becomes negative for $q^2<\Lambda^2_{\overline{\rm MS}}$. To avoid the
resulting numerical ambiguities we first moved this singularity to $q^2=0$
and used a modified form of $\alpha_s(q)$ in the actual numerical
calculations; namely,
\begin{eqnarray}
\alpha_s(q) &=&
\displaystyle\frac{4\pi}{b_0\ln\left(\displaystyle
\frac{q^2}{\Lambda^2_{\overline{\rm MS}}}+1\right)}\,.
\label{al2}
\end{eqnarray}
The value of $\Lambda_{\overline{\rm MS}}$ was taken to be $200$ MeV
and $250$ MeV in the numerical calculations, which is within the experimental
range $\Lambda_{\overline{\rm MS}}=195 + 65 -50 $ MeV~\cite{19}.

To explore the sensitivity of our results to the location of the
Landau singularity in $\alpha_s(q)$ we replaced the expression,
Eq.~(\ref{al2}), by
\begin{eqnarray}
\alpha_s(q) &=&
\displaystyle\frac{4\pi}{b_0\ln\left(\displaystyle
\frac{q^2}{\Lambda^2_{\overline{\rm MS}}}+\lambda^2 \right)}\,
\label{al4}
\end{eqnarray}
and varied $\lambda^2$. The results for the $S$- wave hyperfine splitting
were not sensitive to $\lambda^2$ and only for large $\lambda^2$ did
$\Delta M_p$ significantly decrease. To fit $\Delta M_p$ to the measured
value required $\lambda^2$ quite large, about 16, clearly out of the
perturbative region.

Our numerical results for the $^3S_1-^1S_0$ and
$^3P_J- ^1P_1$ splittings, where $^3P_J$
denotes the center of gravity of the
$^3P_{0,1,2}$ states, for the three potentials\cite{30,31,8}
are presented in
Tables I, II, III, respectively. For comparison we have also included
the results for the $2S$ and $2P$ states.

\eject

\begin{table}[h]
\tcaption{
The hyperfine spin splittings in MeV predicted
 using the Cornell potential~\cite{30}
 }\label{tab:1}
\begin{center}
\small
  \begin{tabular}{lrrrrrr} \hline\hline
  & \multicolumn{2}{c}{~~$c\bar{c}$      }
  & \multicolumn{2}{c}{~~$b\bar{b}$      }
  & \multicolumn{2}{c}{~~$b\bar{c}$      }\\
  ~~~$\Lambda_{\overline{\rm MS}}$ (MeV)
 & ~~~~  200~ & ~~~~  250~
 & ~~~~  200~ & ~~~~  250~
 & ~~~~  200~ & ~~~~  250~
\\ \hline
 $E(1^3S_1)-E(1^1S_0)$ & 117.1 & 128.3 &  97.7 & 104.3 &  67.0 &  71.4\\
 $E(2^3S_1)-E(2^1S_0)$ &  75.3 &  82.5 &  39.6 &  42.2 &  37.7 &  40.3\\
 $E(1^3P_J)-E(1^1P_1)$ &  -4.8 &  -7.0 &  -3.0 &  -4.1 &  -4.0 &  -5.6\\
 $E(2^3P_J)-E(2^1P_1)$ &  -3.6 &  -5.2 &  -2.1 &  -2.9 &  -3.0 &  -4.2\\
\hline\hline
\end{tabular}
\end{center}
\end{table}

\begin{table}[h]
\tcaption{The hyperfine spin splittings  in MeV predicted
using the Logarithmic potential~\cite{31}
 }\label{tab:2}
\begin{center}
\small
  \begin{tabular}{lrrrrrr} \hline\hline
  & \multicolumn{2}{c}{~~$c\bar{c}$      }
  & \multicolumn{2}{c}{~~$b\bar{b}$      }
  & \multicolumn{2}{c}{~~$b\bar{c}$      }\\
  ~~~$\Lambda_{\overline{\rm MS}}$ (MeV)
 & ~~~~  200~ & ~~~~  250~
 & ~~~~  200~ & ~~~~  250~
 & ~~~~  200~ & ~~~~  250~
\\ \hline
 $E(1^3S_1)-E(1^1S_0)$ & 106.1 & 117.1 &  36.0 &  38.2 &  40.6 &  42.6\\
 $E(2^3S_1)-E(2^1S_0)$ &  54.2 &  59.7 &  18.7 &  19.8 &  21.2 &  22.3\\
 $E(1^3P_J)-E(1^1P_1)$ &  -5.4 &  -7.8 &  -3.4 &  -4.5 &  -4.5 &  -6.4\\
 $E(2^3P_J)-E(2^1P_1)$ &  -2.6 &  -3.8 &  -2.1 &  -2.8 &  -2.8 &  -4.0\\
\hline\hline
\end{tabular}
\end{center}
\end{table}
\begin{table}[h]
\tcaption{The hyperfine spin splittings  in MeV predicted
using the Improved-QCD motivated potential~\cite{8}
 }\label{tab:3}
\begin{center}
\small
  \begin{tabular}{lrrrrrr} \hline\hline
  & \multicolumn{2}{c}{~~$c\bar{c}$      }
  & \multicolumn{2}{c}{~~$b\bar{b}$      }
  & \multicolumn{2}{c}{~~$b\bar{c}$      }\\
  ~~~$\Lambda_{\overline{\rm MS}}$ (MeV)
 & ~~~~  200~ & ~~~~  250~
 & ~~~~  200~ & ~~~~  250~
 & ~~~~  200~ & ~~~~  250~
\\ \hline
 $E(1^3S_1)-E(1^1S_0)$ & 107.9 & 119.1 &  44.6 &  47.6 &  43.4 &  45.7\\
 $E(2^3S_1)-E(2^1S_0)$ &  68.5 &  75.6 &  20.9 &  22.4 &  25.2 &  26.7\\
 $E(1^3P_J)-E(1^1P_1)$ &  -4.6 &  -6.7 &  -2.7 &  -3.7 &  -3.7 &  -5.3\\
 $E(2^3P_J)-E(2^1P_1)$ &  -3.4 &  -5.0 &  -1.8 &  -2.5 &  -2.6 &  -3.8\\
\hline\hline
\end{tabular}
\end{center}
\end{table}

The main features of these results can be summarized
as follows:
\begin{itemize}
\item The results are $\mu$-independent, as they must be.
\item The calculated energy difference between the $J/\Psi$ and the $\eta_c$
  mesons is quite close to the experimental value~\cite{19}
  for all three potentials.
\item For each of  these three potentials  we predict the energy difference
between $\Psi'$ and $\eta_c'$ to lie within the range $55-80$ MeV.
\item  For the $b\bar{b}$ system  there are significant discrepancies
between the Cornell model
and the other two models for the S-states. Since the Cornell model, with
the parameters fit to the $c\bar{c}$ spectrum, does not predict
the $b\bar{b}$ spectrum very well, the
results calculated in the other two models are probably better predictions
for the energy difference between the $\Upsilon(1S)$ and the $\eta_b$
($35-50$ MeV) and between the $\Upsilon(2S)$ and the $\eta_b'$ (20 MeV).
\item The predicted energy difference between $B_c^*$ and $B_c$ meson
is in the range $40-70$ MeV from all three of these models, which is
consistent with previous results~\cite{14}.
\item The calculated value of $\Delta M_p \equiv E(1^3P_J)-E(1^1P_1)$
for the charmonium $1P$ states is in the range of $-4$ to $-6$ MeV, which
has the same sign but is several times larger than the experimental value
of $-0.9\pm 0.2$ MeV~\cite{19}. This is not surprising since there are several
other contributions to $\Delta M_p$ which estimates indicate are comparable
in magnitude to the contribution coming from the hyperfine spin--spin
interaction, $H_{ss}$. In fact, it is surprising that the prediction from
only the one--loop spin--dependent potential is quite close to the experimental
data.
\end{itemize}

To summarize, the hyperfine spin splittings in the $c\bar{c}$,
$b\bar{b}$, and $b\bar{c}$ system  were calculated~\cite{-1}
using the RGE improved perturbative
spin--spin potential~\cite{0}. The results for the hyperfine splittings
of the $S$-wave states agree with the $J/\Psi-\eta_c$ measured
splitting~\cite{19} and the prediction for $\Upsilon-\eta_b$  splitting is
reasonable.
However, the contribution to $\Delta M_p \equiv
E(^3P_J)-E(^1P_1)$  for the charmonium $P$-wave states is somewhat
larger than the experimental data~\cite{19}, although it agrees in sign.
That is, after summing up the leading logarithmic terms and including
them in the perturbation calculations, the agreement with the data is not
as good as the one--loop calculations~\cite{8,9,10,11}.
But there are several additional  contributions that are  possibly
comparable in magnitude. These include the following:

\begin{itemize}
\item{The contributions of the spin--orbit and and tensor potentials in
the second order of perturbation theory: These contributions to
$\Delta M_p$ only cancel to first order in perturbation theory.
However, according to the power counting rules~\cite{15},
 the spin--orbit and tensor potential potentials
shift the energies of the $P$-wave states by an amount of order
$mv^4$ in first order, which indeed cancel in $\Delta M_p$, but they
do make a contribution to $\Delta M_p$ of order $mv^6$ in the second
order of perturbation theory. This estimate is several MeV for the
$P$-wave charmonium states, and therefore should not be ignored.}
\item{
Higher dimensional operators: Unlike the dimension--six operators,
these give non-zero contributions to $\Delta M_p$ even at tree level.
Compared to the one--loop contribution, these are suppressed by $v^2$ but
enhanced by $\alpha_s^{-1}$ and $v^2/\alpha_s\sim 1$ in charmonium.
}
\item{
The color-octet $S$-wave component in $P$-wave quarkonia states~\cite{15}:
This component of the wavefunction receives a tree--level contribution from
the local term $\delta^3({\rm\bf r})$ in the spin--spin potential. This
contribution too could be of order $v^2/\alpha_s\sim 1$ compared to what
has been calculated.
}
\item{
Next--to--leading order contributions from the two--loop
potential: These are suppressed by order $\alpha_s$, but since $\alpha_s$
is not a very small quantity in  charmonium, one cannot dismiss the
possibility that this contribution could be significant.
}
\end{itemize}
Before comparing with the experimental value of $\Delta M_p$ in charmonium,
which is only about $1$ MeV, all the above contributions should be included
since they are possibly comparable in magnitude. In the $b\bar{b}$ case
these effects are less important and one can expect the perturbative
calculations the $b\bar{b}$ system to be more reliable, although
$\Delta M_p$ is smaller, also.

 This research was done in collaboration with Yu-Qi Chen and Yu-Ping Kuang
and is presented in greater detail in references [1] and [2].
It was supported in part by the U.S. National Science
Foundation under Grant No. PHYS89-04035, the National Science Foundation
of China, the Fundamental Research Foundation of Tsinghua University, and
the U.S. Department of Energy, Division of High Energy Physics, under
Grant No. DE-FG02-91-ER40684.


\end{document}
\end